\documentclass[11pt,twoside]{article}
\usepackage[pdftex]{graphicx}
\usepackage{amsmath}
\usepackage{amssymb}
\usepackage{cite}
\begin{document}
\newcommand{\pst}{\hspace*{1.5em}}


\newcommand{\be}{\begin{equation}}
\newcommand{\ee}{\end{equation}}
\newcommand{\bm}{\boldmath}
\newcommand{\ds}{\displaystyle}
\newcommand{\bea}{\begin{eqnarray}}
\newcommand{\eea}{\end{eqnarray}}
\newcommand{\ba}{\begin{array}}
\newcommand{\ea}{\end{array}}
\newcommand{\arcsinh}{\mathop{\rm arcsinh}\nolimits}
\newcommand{\arctanh}{\mathop{\rm arctanh}\nolimits}
\newcommand{\bc}{\begin{center}}
\newcommand{\ec}{\end{center}}

\thispagestyle{plain}

\label{sh}


\begin{center} {\Large \bf
\begin{tabular}{c}
Density matrix form of Gross-Pitaevskii equation\end{tabular}
}
\end{center}

\bigskip

\bigskip

\begin{center} {\bf
V. N. Chernega, O. V. Man'ko$^*$, V. I. Man'ko }
\end{center}

\medskip

\begin{center}
{\it P.~N.~Lebedev Physical Institute, Russian Academy of Sciences\\
Leninskii Prospect 53, Moscow 119991, Russia}
\end{center}

\smallskip

\smallskip

$^*$Corresponding author e-mail:~~~omanko@sci.lebedev.ru

\begin{abstract}
\noindent We consider the generalized pure state density matrix which depends on different time moments. The evolution equation for this density matrix is obtained in case where the density matrix corresponds to the solutions of Gross-Pitaevskii equation.
\end{abstract}

\medskip

\noindent{\bf Keywords:} wave function, density matrix, von Neumann equation, nonlinear Schr\"odinger equation.

\section{Introduction}
\pst The wave function $\psi(x,t)$ satisfies the linear Schrodinger equation \cite{Sch26}. The density matrix introduced in \cite{Landau,vonNeuman,vonNeuman1} satisfies the von Neumann equation. There exists nonlinear Schrodinger equation for which potential energy term provides qubic nonlinearity of the equation. The equation was used by Gross \cite{Gross} and Pitaevskii \cite{Pit}. The von Neumann equation for density matrix can be derived from the linear Schrodinger equation. Recently \cite{mamapapa} the generalized density matrix of pure and mixed states was introduced and this matrix depends on two time moments. For equal times the generalized density matrix coincides with the standard density matrix . The equation for this generalized density matrix was discussed in \cite{mamapapa}. The analog of nonlinear von Neumann equation written for Wigner function \cite{Wig32} in case of Gross-Pitaevskii equation was discussed in \cite{RenatoFedeleDeNicolamamapapa}. The aim of our article is to obtain nonlinear analog of Gross-Pitaevskii equation written for generalized density matrix introduced in \cite{mamapapa}.
The paper is organized as follows. In Sec.2 the derivation of von Neumann equation for density matrix is reviewed. In Sec.3 the derivation of nonlinear equation for generalized density matrix corresponding to Gross-Pitaevskii equation is demonstrated. The conclusions and perspectives are given in Sec.4.

\section{Schr\"odinger and von Neumann equations}
The wave function $\psi(x,t)$ of the quantum system satisfies the Schr\"odinger evolution equation
\begin{equation}\label{eq.1}
i\frac{\partial \psi(x,t)}{\partial t}=-\frac{1}{2}\frac{\partial^2\psi(x,t)}{\partial x^2}+U(x)\psi(x,t).
\end{equation}
We assume Planck constant $\hbar=1$ and mass $m=1$. The term $U(x)$ is the real potential energy. To derive von Neumann equation we write the equation for the function $\psi^\ast(x',t)$ which reads:
\begin{equation}\label{eq.2}
-i\frac{\partial \psi^\ast(x',t)}{\partial t}=-\frac{1}{2}\frac{\partial^2\psi^\ast(x',t)}{\partial {x'}^2}+U(x')\psi^\ast(x',t).
\end{equation}
From the system of equations (\ref{eq.1},\ref{eq.2}) it follows the equation for density matrix
\begin{equation}\label{eq.3}
\rho_{\psi}(x,x',t)=\psi(x,t)\psi^\ast(x',t)
\end{equation}
in position representation. The equation reads
\begin{equation}\label{eq.4}
i\frac{\partial\rho_{\psi}(x,x',t)}{\partial t}=-\frac{1}{2}\Large(\frac{\partial^2\rho_{\psi}(x,x',t)}{\partial x^2}-\frac{\partial^2\rho_{\psi}(x,x',t)}{\partial x'^2}\Large)+\large(U(x)-U(x')\large)\rho_{\psi}(x,x',t).
\end{equation}
The same equation (\ref{eq.4}) is valid for  convex sum of pure state density matrices
\begin{equation}\label{eq.5}
\rho(x,x',t)=\sum_n p_n\rho_{\psi_n}(x,x',t),
\end{equation}
where $1\geq p_n\geq0,\quad\sum_n p_n=1.$ For stationary states with wave functions of the factorized form
\begin{equation}\label{eq.6}
\psi_n(x,t)=\phi_n(x)\exp(-iE_n t)
\end{equation}
the density matrix $\rho_{\psi_n}(x,x',t)$ does not depend on time.

In \cite{mamapapa} the generalized density matrix of pure state was introduced
\begin{equation}\label{eq.7}
R_{\psi}(x,x',t,t')=\psi(x,t)\psi^\ast(x',t').
\end{equation}
For equal times $t=t'$ one has equality
\begin{equation}\label{eq.8}
R_{\psi}(x,x',t,t)=\rho(x,,x',t).
\end{equation}
Using the same method of derivation which is used to obtain von Neumann equation from Schr\'odinger equation one can get the equation for the generalized density matrix which reads \cite{mamapapa}
\begin{eqnarray}\label{eq.9}
&&i\frac{\partial R_{\psi}(x,x',t,t')}{\partial t}+i\frac{\partial R_{\psi}(x,x',t,t')}{\partial t'}=\nonumber\\
&&-\frac{1}{2}\Large[\frac{\partial^2 R_{\psi}(x,x',t,t')}{\partial x^2}-\frac{\partial^2 R_{\psi}(x,x',t,t')}{\partial {x'}^2}\Large]+\large[U(x)-U(x')\large] R_{\psi}(x,x',t,t').
\end{eqnarray}
If the times coincide $t=t'$ the above equation provides the von Neumann equation for density matrix $\rho(x,,x',t)$. In view of linearity of Eq.(\ref{eq.9}) the same equation is valid for the matrix
\begin{equation}\label{eq.10}
 R_{\psi}(x,x',t,t')=\sum_n p_n R_{\psi_n}(x,x',t,t').
\end{equation}
Here the generalized density matrix $R(x,x',t,t')$ is convex sum of pure state matrices $R_{\psi_n}(x,x',t,t')$. The generalized density matrix of pure stationary state contains dependence on time, i.e.
\begin{equation}
R_{\psi_n}(x,x',t,t')=\exp[-iE(t-t')]\phi_n(x)\phi_n^\ast(x').
\end{equation}
Thus for stationary states the matrix $R_{\psi_n}(x,x',t,t')$ contains the information corresponding to the both past and future time moments.

\section{Nonlinear Gross-Pitaevskii equation}
The nonlinear Schr\"odinger equation with qubic nonlinearity has the form
\begin{equation}\label{eq.12}
i\frac{\partial\psi(x,t)}{\partial t}=-\frac{1}{2}\frac{\partial^2\psi(x,t)}{\partial x^2}+U(x)\psi(x,t)+g\psi(x,t)|\psi(x,t)|^2.
\end{equation}
Our purpose is to use this equation where $g$ is nonlinearity constant and to derive the corresponding equation for the generalized density matrix $R_{\psi}(x,x',t,t')$.

First, we write the equation for complex conjugate function $\psi^\ast(x',t)$ analogously to the case of nonlinear Schr\"odinger equation
\begin{equation}\label{eq.13}
-i\frac{\partial\psi^\ast(x',t')}{\partial t'}=-\frac{1}{2}\frac{\partial^2\psi^\ast(x',t')}{\partial {x'}^2}+U(x')\psi^\ast(x',t')+g\psi^\ast(x',t')|\psi(x',t')|^2.
\end{equation}
Then we multiply all the terms in Eq.(\ref{eq.12}) by the function $\psi^\ast(x',t')$ and all the terms in Eq.(\ref{eq.13}) by the function $\psi(x,t)$. We get two equations
\begin{equation}\label{eq.14}
i\frac{\partial\psi(x,t)\psi^\ast(x',t')}{\partial t}=-\frac{1}{2}\frac{\partial^2\psi(x,t)\psi^\ast(x',t')}{\partial {x}^2}+U(x)\psi(x,t)\psi^\ast(x',t')+g\psi(x,t)\psi^\ast(x',t')|\psi(x,t)|^2.
\end{equation}
and
\begin{equation}\label{eq.15}
-i\frac{\partial\psi(x,t)\psi^\ast(x',t')}{\partial t'}=-\frac{1}{2}\frac{\partial^2\psi(x,t)\psi^\ast(x',t')}{\partial {x'}^2}+U(x')\psi(x,t)\psi^\ast(x',t')+g\psi(x,t)\psi^\ast(x',t')|\psi(x',t')|^2.
\end{equation}
Taking the difference of equations (\ref{eq.12}) and (\ref{eq.15}) we get the equation
\begin{eqnarray}\label{eq.16}
&&i\frac{\partial R_\psi(x,x',t,t')}{\partial t}+i\frac{\partial R_\psi(x,x',t,t')}{\partial t'}=-\frac{1}{2}\Large[\frac{\partial^2 R_\psi(x,x',t,t')}{\partial x^2}-\frac{\partial^2 R_\psi(x,x',t,t')}{\partial {x'}^2}\Large] \nonumber\\
&&+\large[U(x)-U(x')\large]R_\psi(x,x',t,t')+g R_\psi(x,x',t,t')\large[R_\psi(x,x,t,t)-R_\psi(x',x',t',t')\large].
\end{eqnarray}

The obtained equation is nonlinear and nonlocal equation for the generalized density matrix corresponding to the Gross-Pitaevskii equation.

\section*{Conclusion}
To resume we point out the main results of our work. We reviewed the derivation of von Neumann equation for density matrix from the Schr\"odinger equation. Using the method to provide this derivation we applied it to derive the equation for generalized density matrix which contains time-dependence even for stationary states. We extended the derivation to the case of nonlinear cubic Schr\"odinger equation. The obtained equation gives also the equation for standard density matrix in the case of taking into account the nonlinearity. The forms of the obtained equation in terms of Wigner-Weyl function, Husimi function and tomographic probability distribution will be given in future publications.


\end{document}